\documentclass[pra,aps,twocolumn, reprint, amsmath,amssymb,showpacs,superscriptaddress]{revtex4}

\usepackage{graphicx}
\usepackage{dcolumn}
\usepackage{bm}
\usepackage{graphicx}
\usepackage{epsfig}
\usepackage{epsf}
\usepackage{amssymb}
\usepackage{amsmath}
\usepackage{amsthm}
\usepackage{amssymb}
\usepackage{multirow}
\usepackage{cases}
\usepackage{enumitem}
\usepackage[colorlinks=true,linkcolor=blue,citecolor=blue,pdfauthor={ },pdftitle={ },pdfsubject={ },pdfkeywords={ }]{hyperref}
\usepackage{pgfplots}
\usepackage{dcolumn}
\newcolumntype{d}[1]{D{.}{.}{#1}}

\theoremstyle{definition}

\newtheorem{lemma}{Lemma}
\usepackage{algorithm}
\usepackage{algpseudocode}
\usepackage{algpascal}
\alglanguage{pseudocode}

\usepackage{tikz}
\usepackage{pgfplots}
\usepackage{tikz-3dplot}

\begin{document}

\title{Compilation  of selective pulse network on liquid-state nuclear magnetic resonance system}

\author{Jun Li}
\email{lijunwu@mail.ustc.edu.cn}
\affiliation{Beijing Computational Science Research Center, Beijing 100084, China}
\affiliation{Hefei National Laboratory for Physical Sciences at Microscale and Department of Modern Physics, University of Science and Technology of China, Hefei, Anhui 230026, China}

\author{Jiangyu Cui}
\affiliation{Hefei National Laboratory for Physical Sciences at Microscale and Department of Modern Physics, University of Science and Technology of China, Hefei, Anhui 230026, China}

\author{Raymond Laflamme}
\affiliation{Institute for Quantum Computing and Department of Physics and Astronomy, University of Waterloo, Waterloo, Ontario N2L 3G1, Canada}
\affiliation{Perimeter Institute for Theoretical Physics, Waterloo, Ontario N2L 2Y5, Canada}
\affiliation{Canadian Institute for Advanced Research, Toronto, Ontario M5G 1Z8, Canada}

\author{Xinhua Peng}
\email{xhpeng@ustc.edu.cn}
\affiliation{Hefei National Laboratory for Physical Sciences at Microscale and Department of Modern Physics, University of Science and Technology of China, Hefei, Anhui 230026, China}
\affiliation{Synergetic Innovation Center of Quantum Information $\&$ Quantum Physics,
University of Science and Technology of China, Hefei, Anhui 230026, China}

\begin{abstract}
In creating a large-scale quantum information processor, the ability to construct control pulses for implementing an arbitrary quantum circuit in a scalable manner is an important requirement. For liquid-state nuclear magnetic resonance (NMR)  quantum computing,   a  circuit is generally realized  through a sequence of selective soft pulses, in which various control imperfections exist and are to be  corrected. In this work, we present a comprehensive analysis of the errors arisen in a selective pulse network by using the  zeroth and first order average Hamiltonian theory.   Effective correction rules are derived for adjusting  important  pulse parameters such as irradiation frequencies, rotational angles and transmission phases of the selective pulses to increase the control fidelity. Simulations show that applying our compilation procedure  for a given circuit is  efficient and can greatly reduce the error accumulation.  
\end{abstract}


\maketitle

\section{Introduction}

The prospect of building quantum information processors that probably  outperform their classical counterparts has sparked enormous interest in this field of research \cite{NC00}. However, the development of actual quantum computers remains in its infancy that lots of  current relevant experimental works are still proof-of-principle  demonstrations and are carried on relatively small number of qubits  \cite{VSBYSC01}. Bringing the great  potential of quantum computers to reality requires  precise and universal control over the degrees of freedom that are used as qubits \cite{DP10}. The challenges  lie not only on robust control against noises from  the environment, but also on how to reduce  the operational errors due to imperfections in the control procedure. The task of constructing effective pulse controls is demanding for large quantum systems because of  their huge Hilbert state space. It is thus natural to take scalable pulse control synthesization as one of the central problems in the  quantum control research field. 

Liquid-state NMR  offers  an excellent testbed for benchmarking various quantum algorithms \cite{CVZLL98,JM98,VSBYSC01,WLZLFP11} and developing sophisticated quantum control techniques \cite{KRKSG05,BCR10,LZBGS10,ZLSBG11}. Finding  appropriate control fields that can accomplish the target unitary propagator is crucial for these  experiments. Employment of optimal control theory to  solve this  problem proves to be a great success. In small-size systems, through gradient-based numerical search   high accuracy solutions (e.g., well beyond an error correction threshold around $10^{-4}$ per gate in the case of depolaring noise) can be sought. However, as the method involves in full simulation of the system evolution, it is intrinsically inefficient.  The alternative approach to brute force optimization, is to implement the quantum gates through soft pulses that are of predefined shapes like rectangular or Gaussian wave \cite{Ryan08}.  For example, a rotational gate on an assigned qubit  can be realized by a rotating Gaussian that is on resonance, but just approximately. The imperfections have to be figured out and taken care of. Actually they are  captured by lower order average Hamiltonian theory, thus avoiding simulation of full range of the system dynamics. As to an entire quantum circuit and its corresponding soft pulse network, a systematic and scalable compilation programme  to reduce the imperfections  was built \cite{Ryan08}. The feasibility of this method was confirmed in a number of experiments \cite{KLMT00,VSBYSC01}, and complicated dynamic control has been achieved for a system with up to 12  qubits \cite{N06}.   

The studies on the pulse network compiler technique in the above mentioned works are  limited to corrections of errors that arise from zeroth order average Hamiltonian \cite{Ryan08}. This restricts the kind of operations that can be dealt with in this framework. For instance, Bloch-Siegert shift as a first order effect causes phase errors (which can be moved freely within the pulse network) for single-qubit rotations, but would induce off-resonance errors for  operations that rotate several qubits simultaneously. Therefore, in this paper  we take into account the first order error terms so as to increase the control fidelity to a higher level. With the application of average Hamiltonian theory, we are able to derive a set of general correction rules for  adjusting the selective pulse shapes. We also make numerical tests  on  concrete NMR molecules, including the   control tasks of multi-frequency excitation, multiple-qubit rotation and systematic circuit compilation, to demonstrate the effectiveness of our correction rules.


\section{NMR}
\label{NMR}
To start, we give a brief description of the basics of liquid-state NMR system. Interested readers are refered to standard textbooks such as \cite{Levitt08,Ernst08}  for more details. The liquid NMR sample we are considering consists of an ensemble of  $n$ non-magnetically equivalent spin-1/2 nuclei. The $2\times 2$ identity $\bm{1}$ and the spin operators 
\[ I_x =\frac{1}{2} \left(\begin{array}{cc}
  0 & 1 \\ 
  1 & 0
 \end{array}\right),
 I_y =\frac{1}{2} \left(\begin{array}{cc}
  0 & -i \\ 
  i & 0
 \end{array}\right),
 I_z =\frac{1}{2} \left(\begin{array}{cc}
  1 & 0 \\ 
  0 & -1
 \end{array}\right),
 \]
and their direct products, are introduced to form a basis of the system's state space. On occasions we shall use  the shorthand notation $I_\phi = \cos(\phi) I_x + \sin(\phi)I_y$, which satisfies the commuting relations: (i) $[I_z, I_{\phi}] = i  I_{\phi + \pi/2}$ and (ii) $[I_{\phi_1}, I_{\phi_2}] = i\sin(\phi_2 - \phi_1) I_z$. The sample is placed in a large static magnetic field (strength $B_0$) with its direction, by convention, being defined as the $\hat z$ axis. 
This magnetic field splits the energy levels of the spin states aligned with and against it, giving a  Hamiltonian in Zeeman terms, that is,
\begin{equation}
\label{Omega}
H_\Omega = \sum\limits_{k=1}^n {\gamma_k B_0 (1+\delta_k) I_z^k} = \sum\limits_{k=1}^n {\Omega_k I_z^k},
\end{equation}
where $\gamma_k$, $\delta_k$ and $\Omega_k/2\pi$ are the gyromagnetic ratio, chemical shift and Larmor frequency of the $k$-th spin respectively. Each qubit is associated to a value $k$ and addressable through a pulse with a frequency corresponding to its  chemical shift.
For nuclear spins in molecules that are rapidly tumbling, their effective interaction mechanism  are the isotropic part of the indirect electron-mediated interaction (also called $J$ coupling) through chemical bonds.
If the coupling strengths are much smaller than the differences of the resonant frequencies, the system is then considered to be weakly coupled, and the spin-spin Hamiltonian simply takes the following  form
\begin{equation}
\label{J}
H_{J} = \sum\limits_{k < j} {2\pi  {J_{kj}}I_z^k I_z^j},
\end{equation}
where $J_{kj}$ is the coupling strength between spins $j$ and $k$. External radiofrequency (rf) field $\bm{B}(t) = B(t)\cos \phi(t) \hat x + B(t) \sin \phi(t) \hat y$ is exerted upon the sample, providing the control Hamiltonian 
\begin{equation}
{H_C}(t) = \sum\limits_{k = 1}^n {\gamma_k B(t)\left( \cos \phi (t) I_x^k +  \sin \phi (t) I_y^k \right)}.
\end{equation}
An arbitrary unitary operation can be implemented through a control pulse $H_C(t)$ combined with the system Hamiltonian $H_\Omega + H_J$.

One  important type of rf excitation is the frequency selective pulse. It is designed to excite spins over a
limited frequency region, while minimizing influences to spins that are outside this region. A selective pulse can be thought of as a rotating pulse shape that is on-resonance with the nuclei to be excited. 
To judge the selectivity of a given pulse shape,  it is routine to examine its excitation profile, i.e., dependence of the amount of excitation over the resonance offset. A high quality excitation profile should satisfy that: (i) within the excitation region, full excitation is achieved and (ii) outside the excitation region, excitation is suppressed.   Selective pulse shape design has been extensively studied in the NMR literature and various  optimization methods were proposed \cite{Warren84,GF91,Freeman91,VG04}.  The problem is in principle nonlinear, making it rather difficult to devise a  universal method that applies to all circumstances. However, in the case of small-tip-angle excitations (i.e., linear regime), there does exist a simple description that exploits the close relationship between a rf envelope's excitation profile and its Fourier transform. This linearized treatment results in a number of general and instructive rules in selective pulse shape design. For instance, it suggests that a rf envelope's selectivity increases for longer pulse length  and softer pulse energy. 
The finite-length duration of a pulse also affects its excitation pattern. In the remainder of this section, we present an   analysis of  Gaussian pulse shape and how to choose appropriate parameters  to increase its selectivity. 

\emph{Gaussian selective pulse}.
Let $G(t)$ be an ideal normalized Gaussian envelope that extends to  $\pm \infty$:
\begin{equation}
G(t) = \frac{1}{\displaystyle \sigma \sqrt{2\pi}  } e^{-\frac{t^2}{2\sigma^2}}, 
\end{equation}
here $\sigma$ is the variance parameter.
In the problem of small flip angle $\theta$ excitation, we use a selective pulse $u(t) = G(t) \theta/2\pi $ which when on resonance would generate the right angle of rotation.
For a spin precessing at frequency $\omega$ and initially polarized at the north pole of the Bloch sphere, the $\hat z$ component of the magnetization of the final state, by zeroth order average Hamiltonian theory, is
\begin{equation}
M_z (\omega) \approx \cos (\theta \big|{\hat G}(\omega) \big|),
\end{equation}
here ${\hat G (\omega)}$ is the Fourier transform of $G(t)$ and is thus again a Gaussian.  Therefore the excitation profile goes
\begin{align}
M_z (\omega) & \approx \cos (\theta e^{-\sigma^2\omega^2/2})  \nonumber \\
& = 1 - \frac{1}{2!} \theta^2 e^{-\sigma^2\omega^2} + \frac{1}{4!} \theta^4 e^{-2\sigma^2\omega^2} - \cdots. \nonumber
\end{align}
The leading term is a Gaussian with variance  $\sigma^{-1}/\sqrt{2}$. Suppose the suppresion region is $|\omega| \ge \bar \omega$, that is, the amount of excitations in this region is below some level. 
Let $M_z (|\omega| \ge \bar \omega) \le 1 - \epsilon$ where $\epsilon \ll 1$ is some prescribed precision requirement, then there should be 
\begin{equation}
\sigma \ge  \frac{ \sqrt{\ln {\theta^2} - \ln{2\epsilon} }}{\bar \omega} . 
\label{sigma}
\end{equation}
A practical  pulse in experiment is of finite length, and hence is of  truncated form. The Fourier transform of this truncated window is the convolution of the Gaussian  with a Dirichlet kernel, which results in the formation of the spectral main-lobe  with accompanying side-lobes whose peak levels depend on the parameter $\sigma$ and $T$. Therefore, the width of the Gaussian has to be sufficiently long so that the truncation only causes negligible errors.  
Fig. \ref{profile} gives numerical simulations showing how the Gaussian parameters affect the excitation profile.

\pgfplotsset{grid style={dotted,black}}
\begin{figure}[t]
\begin{center}
\begin{tikzpicture}   
  \begin{axis}[
    width=\linewidth,height=0.75\linewidth,major tick length=0.01\linewidth,
    xmin=-2500,xmax=2500,ytick={-1,-0.6,-0.2,0.2,0.6,1.0},ymin=-1,ymax=1,
    xlabel={Frequency offset (Hz)}, ylabel={Longitudinal amplitude},ylabel near ticks,
    legend pos=south east,grid=both,clip=false
  ]
  \addplot [blue!50!black] file {180sigma2.txt};
  \addplot [green!50!black] file {180sigma3.txt};
  \addplot [red] file {180sigma4.txt};
  
   \legend{
   {\footnotesize $T=2\sigma$}, {\footnotesize $T=3\sigma$}, {\footnotesize $T=4\sigma$}
   };
   \end{axis}
\end{tikzpicture}
\end{center}
\caption{(Color online)  Excitation profiles ($\pi$ rotation) for Gaussian pulses with the same variance $\sigma$ but of different lengths $T$: simulation of the amplitude of the $\hat z$  component of the magnetization of a spin as a function of its frequency. The supposed suppression region is $|\omega| \ge \bar\omega = 2\pi \times 1000$Hz, where we require  the error to be bounded by $\epsilon = 0.01$. In the simulation, the value of variance is given by Eq. (\ref{sigma}): $\sigma = 3.96 \times 10^{-4}$. It can be readily inferred from the profiles that $T$ has to be sufficiently long to ensure that within the suppressed region side-lobes would not occur.}
\label{profile}
\end{figure}
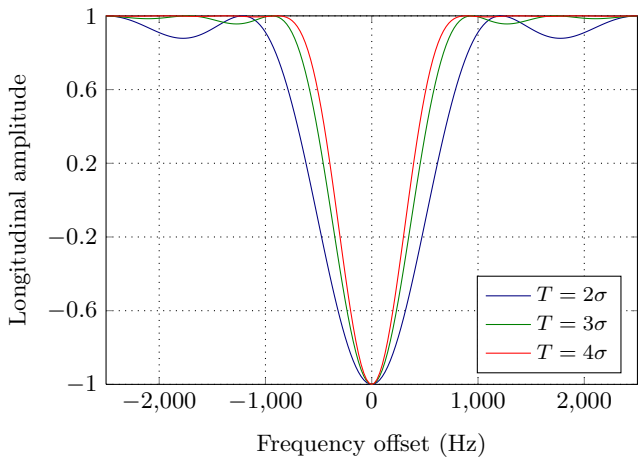


\section{Problem Formulation}
\label{problem}
In the circuit model of quantum computation, a quantum algorithm is implemented through a sequence of single-qubit rotations and two-qubit coupling gates \cite{NC00}. The coupled logic can originate from the natural couplings in the quantum system involved. In solution NMR, the scheme of coupling any designated pair of spins so as to perform a target two-qubit gate can be made efficient by using decoupling and recoupling techniques with ideal $\pi$ pulses \cite{LCYY00}. Therefore, a circuit can always be  decomposed into a train of local single-spin operations separated by free evolutions  between these operations. Let us then, without loss of generality, represent an arbitrary circuit by $\mathcal{C}_{\text{ideal}}$ as follows:
\begin{equation}
\mathcal{C}_{\text{ideal}}=\{(t_p,U_{\text{ideal}}^p)\},
\label{Cideal}
\end{equation}
where $t_p$ is the action time of the $p$-th operation $U_{\text{ideal}}^p$
\begin{equation}
U_{\text{ideal}}^p=\mathop  \bigotimes \limits_{k=1}^n R_{{\varphi _k}}^k({\theta _k}),
\label{Uideal}
\end{equation}
here $ R_{{\varphi _k}}^k({\theta _k}) = e^{-i \theta_k I_{\varphi_k} }$ denotes a  rotational transformation of the $k$-th spin for an angle ${\theta _k}$  ($\theta _k \in [0, \pi)$) along  axis ${\varphi _k}$ ($\varphi_k \in [0,2\pi)$) in x-y plane.
Control pulses that are able to generate these unitary operators can be found by numerical searching according to some distance or fidelity
measure. However, the computational costs would soon become unaffordable for larger systems. The relatively easier alternative would be applying frequency selective pulses. As mentioned in the previous section, a selective pulse shape  on resonance with a specific  spin can excite that spin selectively. Thus, to rotate the $k$-th  qubit, we use a selective pulse of the following form
\begin{equation}
u_k(t) = \frac{\theta_k }{{2\pi }} {G}_k (t) \left(\cos (\Omega_k t + \varphi_k), \sin (\Omega_k t + \varphi_k)\right),
\end{equation}
here $t \in [0, T]$ with $T$ denoting the pulse length,  $ G_k$ is some normalized symmetric pulse shape satisfying
\begin{equation}
\int_0^T { G_k(t) dt} = 1.
\end{equation}
To rotate the spins simultaneously, one intuitively  adds the corresponding selective pulses together
\begin{equation}
u(t) = \sum_{k=1}^n {\frac{\theta_k }{{2\pi }} {G}_k (t) \left(\cos (\Omega_k t + \varphi_k), \sin (\Omega_k t + \varphi_k)\right)}.  \nonumber
\end{equation}
Not surprisingly, this control pulse does not exactly implement the wanted  operation  and becomes less effective for increasing number of qubits that are to be excited. It is then necessary to analyze what kinds of error come in and what corrections must be made to the pulse parameters for the goal of improving  control fidelity.

\begin{figure}[t]
\centering
\includegraphics[width=\linewidth]{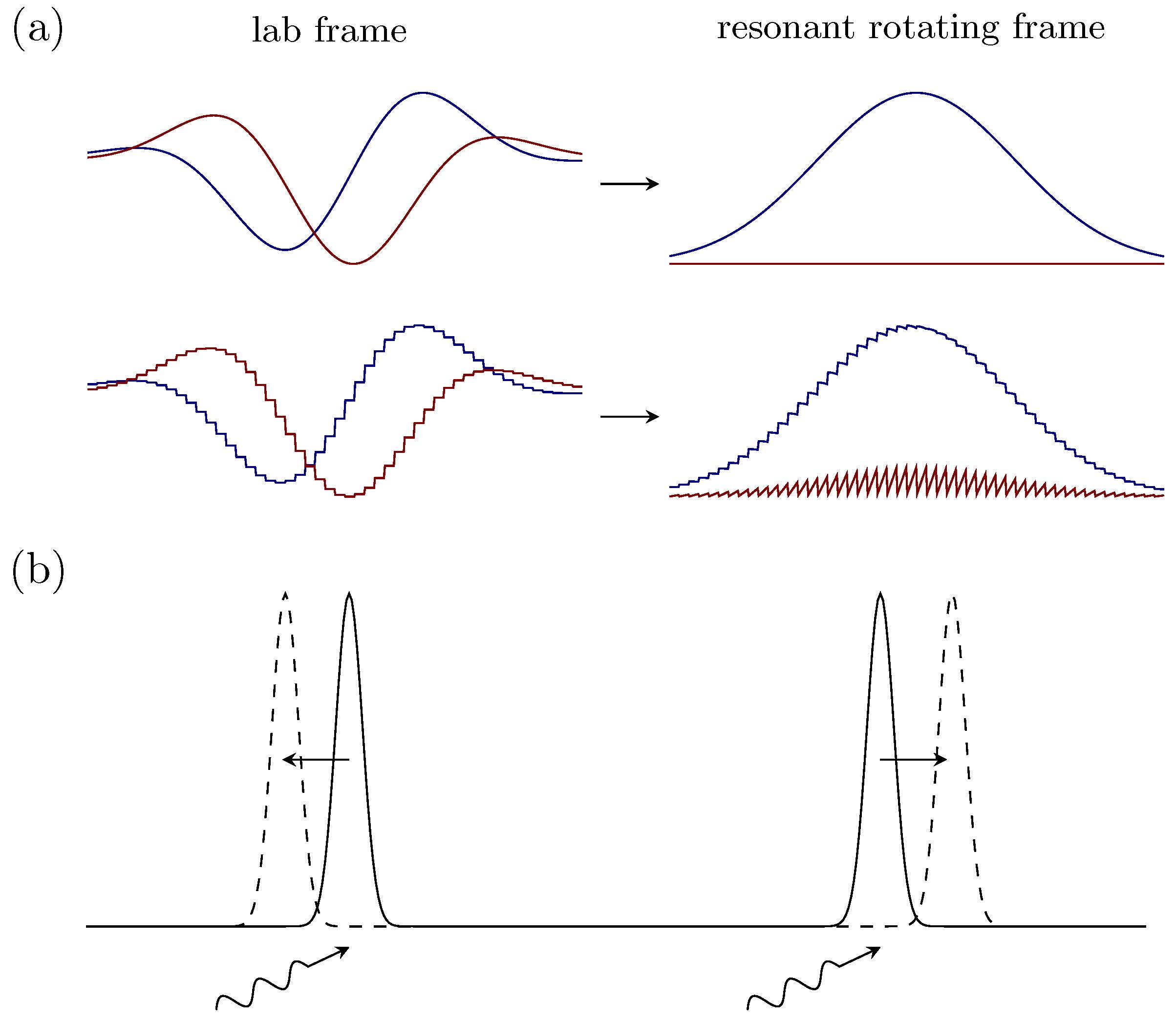}
\caption{(Color online) (a) Errors due to pulse discretization. The left hand side is plotted an ideal smooth single-spin selective pulse and its discretized form (blue: $\hat x$ component field; red: $\hat y$ component  field).  When transformed into the resonant rotating frame, it can be seen that with discretization, the spin would experience a distorted  control field which  rotates the spin about a slightly altered axis.  (b)  Illustration of Bloch-Siegert effect. When spin-selective pulses are applied simultaneously to two spins, the pulse at one spin temporarily shifts the frequency of the another. As a result, the pulse on the other spin, without irradiation frequency correction, will be off-resonance and the resulting rotations deviate significantly from the desired operation.}
\label{error}
\end{figure}

(1) \emph{Choice of $T$ and $\sigma$.} Each  selective pulse envelope $G_k$ should  be of good enough selectivity, which is measured by the ability to excite an assigned resonance without appreciably affecting near neighbours. Given the values of the qubits' precession frequencies, it then gets clear on the configuration of the desired excitation profile.  In mathematical terms, this requires that
\begin{equation}
\left| \int_0^T {G_k(t) e^{-i (\Omega_j-\Omega_k) t)} dt} \right| \approx 0, \quad  \text{for } k\ne j.
\label{Gproperty}
\end{equation}
To meet the condition above, Gaussian  is preferred to rectangle pulse, as its associated excitation profile is better localized in frequency domain \cite{BFFKS84}. Our example presented in the last section is helpful for choosing appropriate values of the Gaussians' variances  and widths.

(2) \emph{Correction of $\theta$ and $\varphi$ due to input pulse discretization.} Although smooth pulses are ideal, the system limitations force us to discretize the pulse into many time slices. There is also the possibility that sometimes we would like the pulse to be of less number of slices so that the time cost of simulating system evolution could be reduced. Fig. \ref{error}(a) suggests that    discretization procedure results in deviations of the rotational angle and  rotational axis. The imperfections here  have to be taken into account.

(3) \emph{Correction of $\Omega$ due to Bloch-Siegert shift.} It has long been known that a selective irradiation on a particular spin will cause an offset-dependent $\hat z$ rotational phase shift to the evolution of  another spin, that is,  the so-called  Bloch-Siegert (BS) shift \cite{BS40}. As a higher order averaging effect, this amounts to shift the resonant frequency of the spins even that are  far outside the excitation region (see Fig. \ref{error}(b)). Eliminating Bloch-Siegert shifts is a frequently encountered  subject and there have been a lot of relavent research works. However, the common methods, such as  using  additional soft pulses to balancing the frequency shifts of the spins \cite{MM93}, or brute-force optimization of the pulse shapes \cite{PRNM91}, are not suited for the present task.  The more convenient approach would be to adjust   the irradiation frequencies  of the pulse $u(t)$ so that each of them can  match the corresponding shifted resonance \cite{SVC00}.

(4) \emph{Pre- and post-error representation.} It is especially important to be aware of that, the propagator (denoted by $U_\text{sim}$) generated by a single-spin selective pulse does not simply equal to the ideal operation. Their actual relation has been exploited in \cite{Ryan08}: to first order approximation, and if $\max_{k<j} |J_{kj} T| \ll 1$, there exists the following general pre- and post-error decomposition scheme
\begin{equation}
U_\text{sim} \approx U_\text{post} \cdot U_\text{ideal} \cdot U_\text{pre}, 
\label{pre-post}
\end{equation}
where $U_\text{pre}$ and $U_\text{post}$ takes the following form
\begin{align}
U_\text{pre} & = \exp\left\{-i\sum\limits_{k=1}^n {\alpha_k^\text{pre} I_z^k} \right\}  \exp\left\{-i\sum\limits_{k<j}^n {\beta_{kj}^\text{pre} 2 I_z^k I_z^j} \right\},  \nonumber \\
U_\text{post} & = \exp\left\{-i\sum\limits_{k=1}^n {\alpha_k^\text{post} I_z^k} \right\}  \exp\left\{-i\sum\limits_{k<j}^n {\beta_{kj}^\text{post} 2 I_z^k I_z^j} \right\}. \nonumber
\end{align}
The parameters $\left\{ \alpha_k^\text{pre}, \alpha_k^\text{post} \right\}$ are phase errors due to spin precession and Bloch-Seigert effect, and $\left\{ \beta_k^\text{pre}, \beta_k^\text{post} \right\}$ are coupled evolution  errors. In \cite{Ryan08} there is described an efficient procedure of numerically seeking for these error terms, which are then compensated by using a general pulse compiler programme.  Although for multi-spin selective excitation the representation Eq. (\ref{pre-post}) still holds (as we will see in the next section), determining the phase errors by numerical optimization would not be a good choice now. Experiences show that phase and coupling errors are not  likely to be both eliminated completely with simple general rules.

We now formulate  the central problem to study in this work  in a more precise way.  We are going to use multiple-qubit selective pulse to implement the ideal operation Eq. (\ref{Uideal}). Let the pulse  be of length $T$, and be divided into $M$ pieces, let $\tau=T/M$ denote the length of each time step, for the $m$-th time interval $t \in [(m-1)\tau, m\tau)$, set
\begin{equation}
u[m] =  \sum\limits_{k=1}^n {\frac{\theta'_k }{{2\pi }} {G}_k [m] (\cos (\Omega'_k m\tau + \varphi'_k), \sin (\Omega'_k m\tau + \varphi'_k )) }. \nonumber
\end{equation}
The corresponding control Hamiltonian reads
\begin{equation}
H_\text{rf}[m] = \sum\limits_{k=1}^n \sum\limits_{j=1}^n {\theta'_j {G}_{j} [m] I^k_{\Omega'_j m\tau + \varphi'_j} },  
\label{control}
\end{equation}
Our goal  is then to find    correction rules for  $\theta'_k$, $\Omega'_j$ and $\varphi'_j$ to increase the fidelity of  representation Eq. (\ref{pre-post}).

\section{Theoretical Derivation}
Average Hamiltonian theory has frequently been used to generate an expansion in the effective Hamiltonian. It offers a powerful framework for analyzing and removing unwanted terms in the Hamiltonian by periodic perturbations  without requiring full knowledge of the system dynamics. The formalism  can be described as follows. In the multi-rotating frame 
\begin{equation}
{\rho ^I} = {\exp\left\{i { \sum\limits_{k=1}^n {\Omega_k I_z^k}} t \right\} }  \cdot \rho \cdot  {\exp \left\{ - i{ \sum\limits_{k=1}^n {\Omega_k I_z^k}}t \right\}}, \nonumber
\end{equation}
the system  evolves according to
\begin{equation}
\dot \rho^I = -i [H_J^I + H_\text{rf}^I(t), \rho^I],
\end{equation}
where $H^I_J = H_J$ and $H^I_\text{rf}(t)$, according to expression Eq. (\ref{control}), is a piecewise continuous function, that is, for the $m$-th  time interval $t \in [(m-1)\tau, m\tau)$:
\begin{equation}
H^I_\text{rf}(t) = \sum\limits_{k=1}^n \sum\limits_{j=1}^n{\theta'_j {G}_j [m] I^k_{\Omega'_j m\tau -\Omega_k t + \varphi'_j} }.
\label{rf}
\end{equation}
Let $\mathcal{T}$ denote the Dyson time ordering operation, the time evolution operator in the interaction picture  will  be formally written as
\begin{equation}
U^I  = \mathcal{T} \left( \int_0^T { \exp\left\{ -i \left(H_J + H^I_\text{rf}(t) \right)t  \right\} dt} \right).
\end{equation}
The Magnus expansion \cite{Magnus54}, permits calculation of the effective propagator $\mathcal{U}^I$ for the entire sequence
\begin{equation}
\mathcal{U}^I  =  \exp\left\{ -i \left(\mathcal{H}^{(0)} + \mathcal{H}^{(1)} + \cdots \right)T\right\},
\end{equation}
in which the first two terms are given by
\begin{align}
\mathcal{H}^{(0)} & = \frac{1}{T} \int_0^T { \left( H_J + H^I_\text{rf}(t_1) \right) dt_1}, \nonumber \\
\mathcal{H}^{(1)} & = \frac{-i}{2T} \int_0^T { \int_0^{t_2} {\left[ H_J + H^I_\text{rf}(t_2), H_J + H^I_\text{rf}(t_1) \right]   dt_1} dt_2}.  \nonumber 
\end{align}
Provided that the applied  pulse is sufficiently soft in a suitable sense (e.g., a converging criterion would be $\int_0^T {\left\| H_J + H^I_\text{rf}(t) \right\| dt \ll C}$ for some constant $C$ \cite{BCOR09}), then $\mathcal{H}^{(0)}$ and $\mathcal{H}^{(1)}$ will capture most of the system behaviours. Now we substitute Eq. (\ref{rf}) into the above expressions. The zeroth order averaged Hamiltonian is
\begin{widetext}
\begin{align}
\mathcal{H}^{(0)} 
& = H_J + \frac{1}{T} \sum\limits_{k=1}^n \sum\limits_{j=1}^n \sum\limits_{m=1}^M {\left(  \int_{(m-1)\tau}^{m\tau} { \theta'_j {G}_j [m] \left[ \cos(\Omega'_j m\tau -\Omega_k t + \varphi'_j) I_x^k + \sin(\Omega'_j m\tau -\Omega_k t + \varphi'_j) I_y^k \right] dt } \right)}  \nonumber \\
& = H_J + \frac{1}{T} \sum\limits_{k=1}^n \sum\limits_{j=1}^n \sum\limits_{m=1}^M {\left(\theta'_j {G}_j [m] \tau\frac{ \sin(\Omega_k\tau/2)}{\Omega_k \tau/2}   I^k_{(\Omega'_j-\Omega_k) m\tau + \Omega_k \tau/2 + \varphi'_j}    \right) } \nonumber  
\end{align}
\end{widetext}
Recall that  we have assumed good selectivity of the Gaussian shapes (see Eq. (\ref{Gproperty})), we here thus for the above expression only retain those terms for which $k = j$ as otherwise the operator $I^k_{(\Omega'_j-\Omega_k) m\tau + \Omega_k \tau/2 + \varphi'_j}$ would be fast oscillating such  that the relative summation vanishes. Now, we  set
\begin{equation}
\varphi'_k = \varphi_k - \Omega_k \tau/2 - \Delta \Omega_k T/2.
\label{varphi}
\end{equation}
for the reason that will be seen instantly.
Then, we have
\begin{equation}
\mathcal{H}^{(0)}  \approx H_J + \frac{1}{T} \sum\limits_{k=1}^n   {\theta'_k \frac{ \sin(\Omega_k\tau/2)}{\Omega_k \tau/2} \left(  \zeta_k   I^k_{\varphi_k} + \zeta'_k   I^k_{\varphi_k+\pi/2}  \right) }. \nonumber
\end{equation}
where
\begin{align}
\zeta_k & =  \sum\limits_{m=1}^M { \left( {G}_k [m] \tau  \cos\left(\Delta\Omega_k (m\tau - T/2)\right)  \right)},  \nonumber \\
\zeta'_k & =  \sum\limits_{m=1}^M { \left( {G}_k [m] \tau \sin\left(\Delta\Omega_k (m\tau - T/2)\right) \right)}.  \nonumber
\end{align}
$\zeta_k$ is close to  1 as $\Delta\Omega_k$ is small. $\zeta'_k$ actually vanishes because $G_k$ is symmetric, while $\sin\left(\Delta\Omega_k (m\tau - T/2)\right)$ is antisymmetric with respect to $t=T/2$.
To ensure that the $k$-th spin is rotated by the desired angle, we set
\begin{equation}
\theta'_k  = \theta_k/\zeta_k ,
\label{theta}
\end{equation}
Consequently, we obtain
\begin{equation}
\mathcal{H}^{(0)} \approx H_J + \frac{1}{T} \sum\limits_{k=1}^n {\theta_k I^k_{\varphi_k}}.
\label{H1}
\end{equation}
Next, we calculate $\mathcal{H}^{(1)}$. Obviously that there is
\begin{equation}
\mathcal{H}^{(1)} = \mathcal{H}^{(1)}_1 + \mathcal{H}^{(1)}_2 + \mathcal{H}^{(1)}_3, 
\end{equation}
where
\begin{align}
\mathcal{H}^{(1)}_1  & = \frac{-i}{2T} \int_0^T { \int_0^{t_2} {\left[ H^I_\text{rf}(t_2),  H^I_\text{rf}(t_1) \right]   dt_1} dt_2}, \nonumber \\
\mathcal{H}^{(1)}_2  & = \frac{-i}{2T} \int_0^T { \int_0^{t_2} {\left[ H_J,  H^I_\text{rf}(t_1) \right]   dt_1} dt_2}, \nonumber  \\
\mathcal{H}^{(1)}_3  & = \frac{-i}{2T} \int_0^T { \int_0^{t_2} {\left[ H^I_\text{rf}(t_2), H_J \right]   dt_1} dt_2}.\nonumber
\end{align}
We will only consider the first term, which corresponds to the transient Bloch-Siegert shift. The estimation procedure is put in the appendix. With a  lengthy derivation,  we   got the following expression (see Eq. (\ref{H111}) (\ref{H112}))
\begin{equation}
\mathcal{H}^{(1)}_1 \approx \sum\limits_{k=1}^n {\left({\theta'_{k}}^2 \frac{\sin^2(\Omega_k\tau/2)}{(\Omega_k \tau/2)^2} \eta_k \Delta\Omega_k + \Delta \Omega_k^{\text{BS}}\right) I^k_z}, \nonumber
\end{equation}
with $\eta_k$ determined by
\begin{equation}
\eta_k \approx  \frac{\tau^3}{2T}  \sum\limits_{m_2 =1}^M \sum\limits_{m_1 =1}^{m_2} {\left[ {G}_{k} [m_2]    {G}_{k} [m_1]   (m_1-m_2)  \right]}  \nonumber
\end{equation}
and $\Delta \Omega_k^{\text{BS}}$ denoting the transient Bloch-Siegert shift for the $k$-th spin:
\begin{equation}
\Delta \Omega_k^{\text{BS}} \approx   \sum\limits_{j \ne k}^n {\left( - \frac{\sin^2(\Omega'_k\tau/2)}{(\Omega'_k \tau/2)^2}  \frac{ {\theta'_{j}}^2 \mathbb{E}({G}^2_j) }{2(\Omega'_{j} - \Omega'_{k})} \right)},
\label{BS}
\end{equation}
here the definition of  $\mathbb{E}$ is given in Eq. (\ref{E}). 
To elimintate the Bloch-Siegert shifts in $\mathcal{H}^{(1)}_1$, we have to set the detunings as follows
\begin{equation}
\Delta \Omega_k = -  \frac{1}{\eta_k {\theta'_{k}}^2} \frac{(\Omega_k \tau/2)^2}{\sin^2(\Omega_k\tau/2)} \Delta \Omega_k^{\text{BS}}.
\label{Omega}
\end{equation}
Eventually, we  get  the result
\begin{equation}
\mathcal{U}^I \approx  \exp \left\{ - i H_J  T - iT \sum\limits_{k=1}^n{\kappa_k \Delta \Omega_k^{\text{BS}} I^k_z  } - i \sum\limits_{k=1}^n  {\theta_k   I^k_{\varphi_k}} \right\}, \nonumber 
\end{equation}
here $\kappa_k$ is 1 if $\theta_k = 0$ and is 0 if $\theta_k \ne 0$.  Back to the lab frame, there is
\begin{align}
U_\text{sim} \approx {} & \exp\left\{ -iT { \sum\limits_{k=1}^n {(\Omega_k + \kappa_k \Delta \Omega_k^{\text{BS}}) I_z^k}}  \right\}  \nonumber \\
{} &\exp \left\{ {\frac{{ - iH_J T}}{2}} \right\} U_\text{ideal}  \exp \left\{ {\frac{{ - iH_J T}}{2}} \right\},
\label{sim} 
\end{align}
here we used the Trotter expansion: ${e^{A + B}} \approx {e^{B/2}}{e^A}{e^{B/2}}$ when $\left\| B \right\|$ is much smaller than $\left\| A \right\|$.

\section{Compilation}
In  the last section, in deriving the pre- and post- representation for $U_\text{sim}$ we have fixed Eq. (\ref{varphi}),(\ref{theta}) and (\ref{Omega}), which we regard as to be the   pulse parameter correction rules that we are seeking for. Solving these error compensation equations requires numerical nonlinear  optimization. Nonetheless, reasonable approximations can be made to simplify the rules, e.g., $\sin(\Omega'_k \tau/2)/(\Omega'_k \tau/2) \approx \sin(\Omega_k \tau/2)/(\Omega_k \tau/2) $ as $\Delta\Omega_k$  is small.
Summarily, we state our main results as follows. 
Suppose we are using control pulse of the form Eq. (\ref{control}) to do operation Eq. (\ref{Uideal}), suppose the selective pulse shapes are symmetric Gaussian envelopes with good selectivity,  and the values of the irradiation frequencies, rotational angles and rotational phases are respectively determined according to the following rules:
{\small
\begin{subequations}
\begin{align}
\Omega'_k & = \Omega_k - \frac{\displaystyle \sum\limits_{j \ne k}^n {\left( - \frac{\sin^2(\Omega_k\tau/2)}{(\Omega_k \tau/2)^2}  \frac{(\Omega_j \tau/2)^2}{  \sin(\Omega_j\tau/2)^2} \frac{ \theta_{j}^2 \mathbb{E}({G}^2_j) }{2(\Omega_{j} - \Omega_{k})} \right)}}{\displaystyle \frac{\theta_{k}^2 \tau^3}{2T}  \sum\limits_{m_2 \ge m_1}^M  { \left[ {G}_{k} [m_2]   {G}_{k} [m_1]   (m_1-m_2)  \right]} }, \\
\varphi'_k & = \varphi_k - \Omega_k \tau/2 - \Delta \Omega_k T/2, \\
\theta'_k  & = \frac{\displaystyle \frac{\Omega_k \tau/2}{  \sin(\Omega_k\tau/2)} \theta_k}{\displaystyle \sum\limits_{m=1}^M {\left( G_k[m] \tau \cos(\Delta\Omega_k (m\tau -T/2) )  \right)}} .
\end{align}
\end{subequations}
}
Then to first order dynamics,  there will be 
\begin{equation}
U_\text{sim} \approx U_\text{post} \cdot U_\text{ideal} \cdot U_\text{pre} 
\label{mainformula}
\end{equation}
with the pre-errors and post-errors given by
\begin{subequations}
\begin{align}
\alpha_k^\text{pre} & =   0, \label{pre-alpha} \\
\alpha_k^\text{post} & = \left(\Omega_k + \kappa_k \Delta \Omega_k^{\text{BS}}\right)  T,  \label{post-alpha} \\
\beta_{kj}^\text{pre} & = \pi J_{kj} T/2,  \label{pre-beta} \\
\beta_{kj}^\text{post} & = \pi J_{kj} T/2.  \label{post-beta}  
\end{align}
\end{subequations}
We give a very quick discussion of the correction rules in  the following two cases:

(1) \emph{With state assumptions.} In the selective pulse driven state transfer problem, if the starting state or the ending state is in some known specific state that are commutative with $\hat z$ rotations, then the phase tracking calculations can be simplified.  Examples of such specific states include pseudopure state and maximally mixed state \cite{Ryan08}. Basically, the correction formula for $\phi_k$ will be: (i) if pre-error term $e^{-i \alpha_{\text{pre}}^k I_z^k}$ does not affect starting state, then $\varphi'_k  = \varphi_k - \Omega_k \tau/2 - \Delta \Omega_k T/2 - (\Omega_k + \kappa_k \Delta \Omega_k^{\text{BS}}) T$; (ii)  if post-error term $e^{-i \alpha_{\text{post}}^k I_z^k}$ does not affect ending state, then $\varphi'_k  = \varphi_k - \Omega_k \tau/2 - \Delta \Omega_k T/2$.

(2) \emph{Zeroth order correction.} If we only correct phase and angle parameters, but keep the  frequencies of the Gaussians  unchanged, this can be thought of as  zeroth order correction, in agreement with \cite{Ryan08}. We will compare the performances of zeroth and first order correction for multiple-qubit rotations in the next section.

Now let's see how to compile a selective pulse network. As indicated by  Eq. (\ref{mainformula}), there mainly involves two types of errors, namely $I_z$ and $I_z I_z$ rotations. Apparently, for a complicated network, most  pulses give the right $I_z I_z$ evolution (see Eq. (\ref{pre-beta}), (\ref{post-beta})),  except for a few that are at the very start or end of the network. While the previous compiler method in Ref. \cite{Ryan08} employed a further subprogramme to optimize the delyas between the pulses, in our formulation we are not going to alter the locations of the selective pulses. Although this strategy  sacrifices  some fidelity for sure, the benefits would be making the   compilation activity rather simple on the whole.
As for $I_z$ error terms, the crucial point is that (see \cite{Ryan08,VC04}),  whenever there is a rotation about $\hat z$ say $R_z(\gamma)$:  (i) if it is followed by a period of free evolution, their order can be interchanged; (ii) if it is followed by a transverse rotation $R_\varphi(\theta)$, it can be moved across that rotation according to: $R_\varphi(\theta) R_z(\gamma) = R_z(\gamma) R_{\varphi-\gamma}(\theta) $. Therefore $\hat z$ rotations, whenever encountered, actually need not be executed and can always be moved one step forward till the end of the network. One keeps in track of  $\hat z$ error accumulation process, and by doing so one can readily figure out the right measurements to take after finishing the circuit.
Next section will give a concrete control instance demonstrating how this method works.

\section{Simulations}
In this section, we perform  numerical simulations using our correction procedure on two physical systems, namely (i) iodotrifluroethylene (C$_2$F$_3$I) dissolved in d-chloroform, which contains three $^{19}$F nuclei:
\begin{center}
\includegraphics[width=\linewidth]{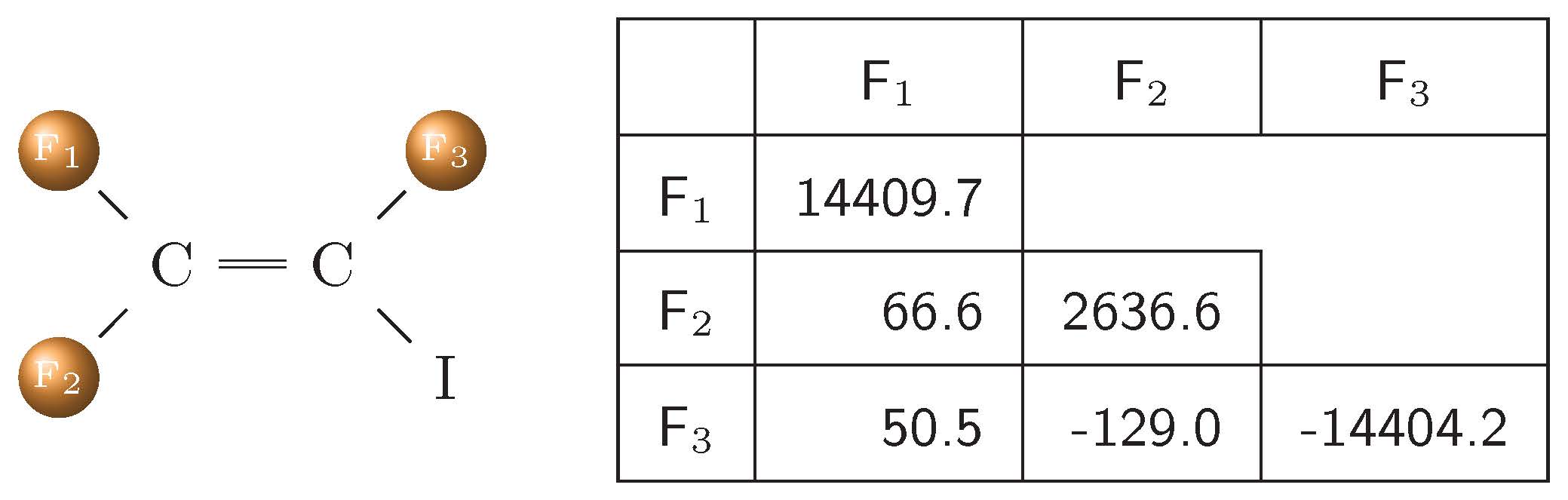}
\end{center}
and (ii)  pre-$^{13}$C-labeled dichlorocyclobutanone derivative dissolved in d6-acetone, which contains seven labeled carbon nuclei and  serves as a seven-qubit system when the $^{1}$H nuclei are  decoupled:
\begin{center}
\includegraphics[width=\linewidth]{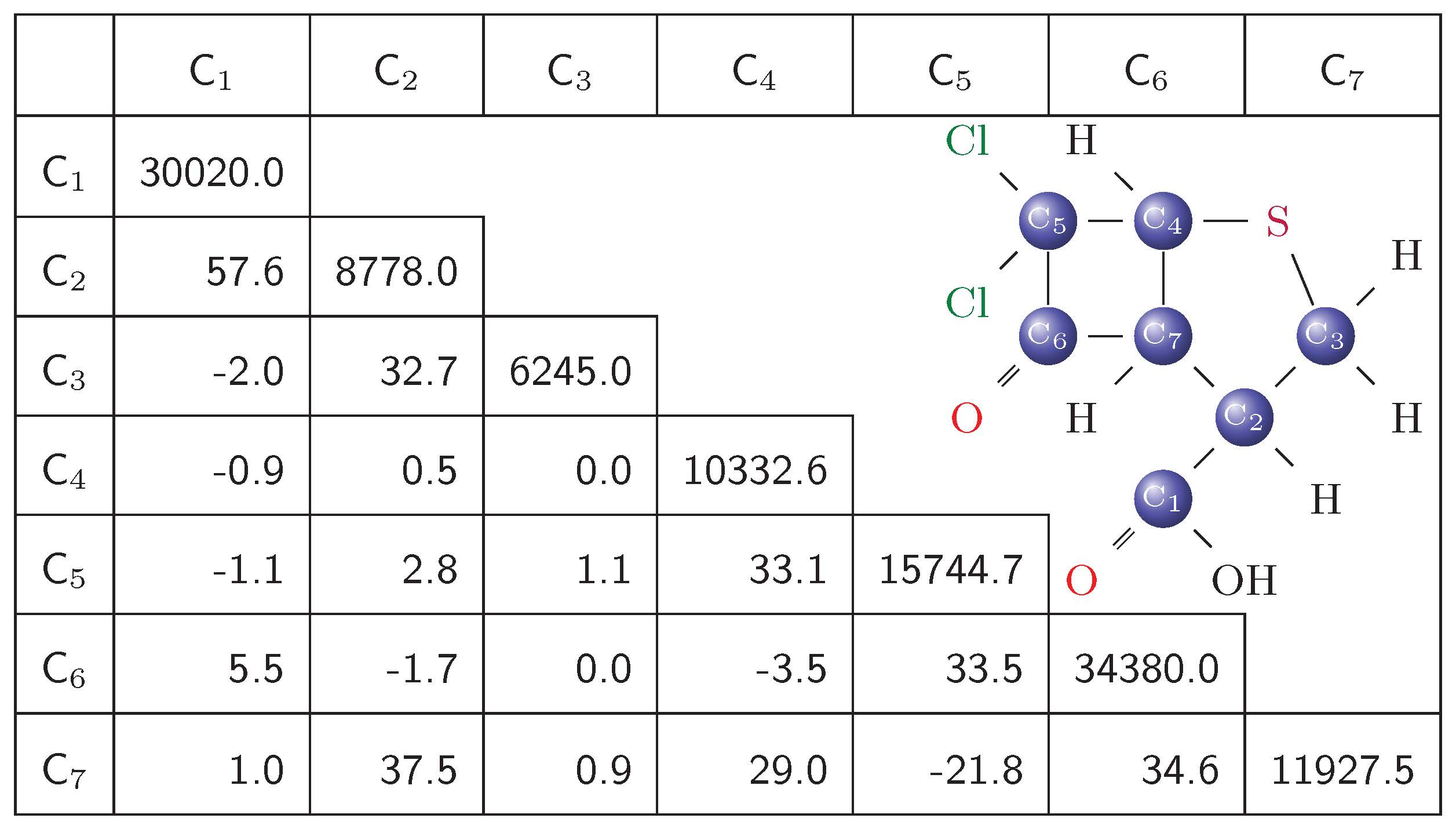}
\end{center}
The system Hamiltonian parameters  (all in Hz) as given in the tables  are extracted from \cite{Wenqiang15} and  \cite{Dawei15} respectively: diagonal elements of the tables give chemical shifts  with respect to the base frequency for $^{19}$F or $^{13}$C transmitters; off-diagonal elements give $J$ coupling terms.

\begin{figure}[t]
\includegraphics[width=\linewidth]{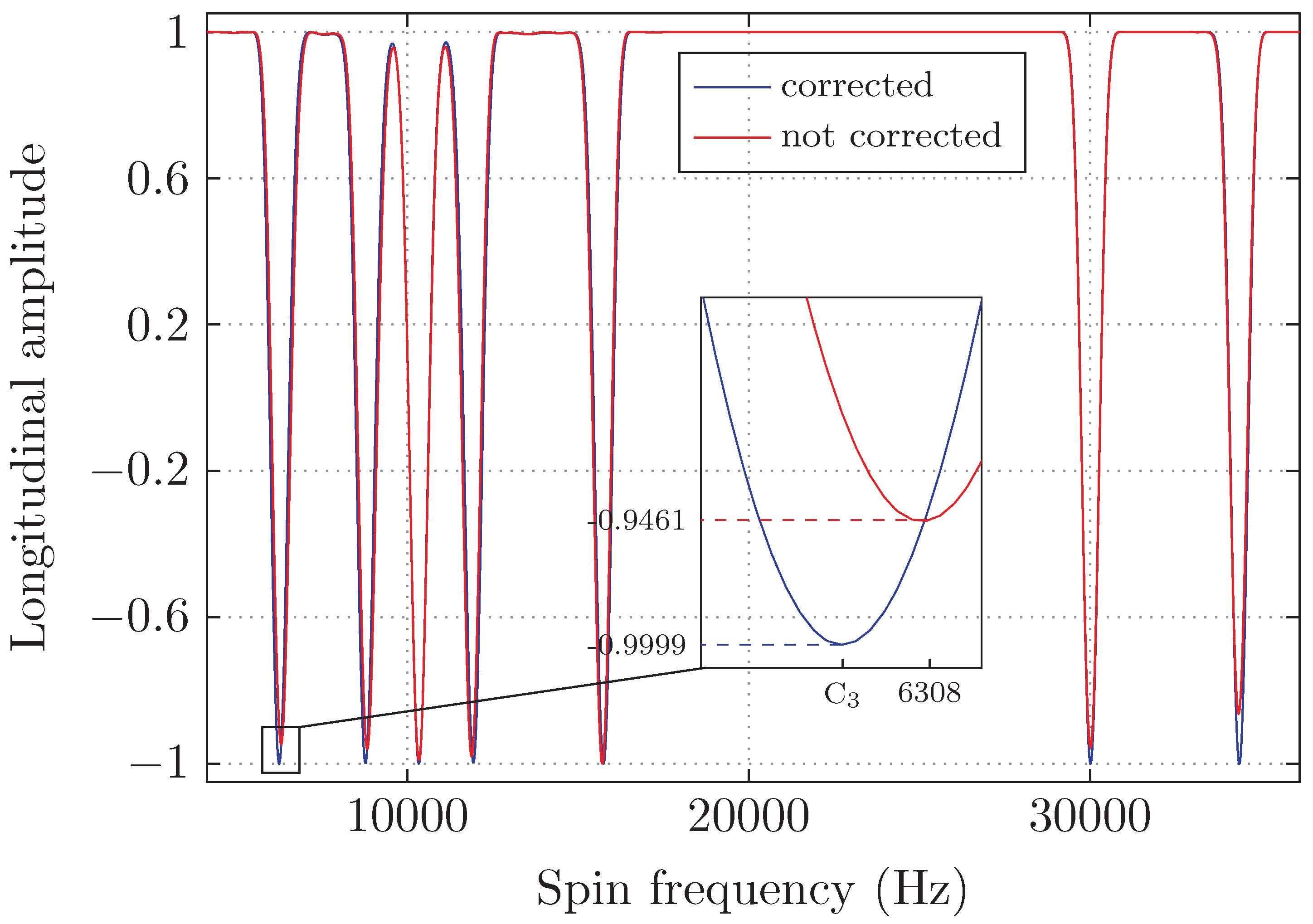}
\caption{(Color online) Simulation of the amplitude of the $\hat z$ component of the magnetization of a spin as a function of its frequency. The spin starts out along $+ \hat z$ and is subject to seven simultaneous Gaussian-shaped pulses ($T=2.2$ ms, $\sigma=T/5$, $\tau =20$ $\mu$s) with carrier frequencies on resonance with dichlorocyclobutane's seven qubits respectively. Shown in the zoomed-up inset plot is a BS shift of 62.53 Hz and unideal inversion, which then is perfectly corrected with parameter calibration. The data of the inverted magnetization   at the seven  sites are  (from left to right in order): -0.9999,-0.9975,-0.9997,-0.9997,-0.9963,-0.9993,-0.9991.}
\label{7qubitProfile}
\end{figure}

\begin{figure}[t]
\includegraphics[width=\linewidth]{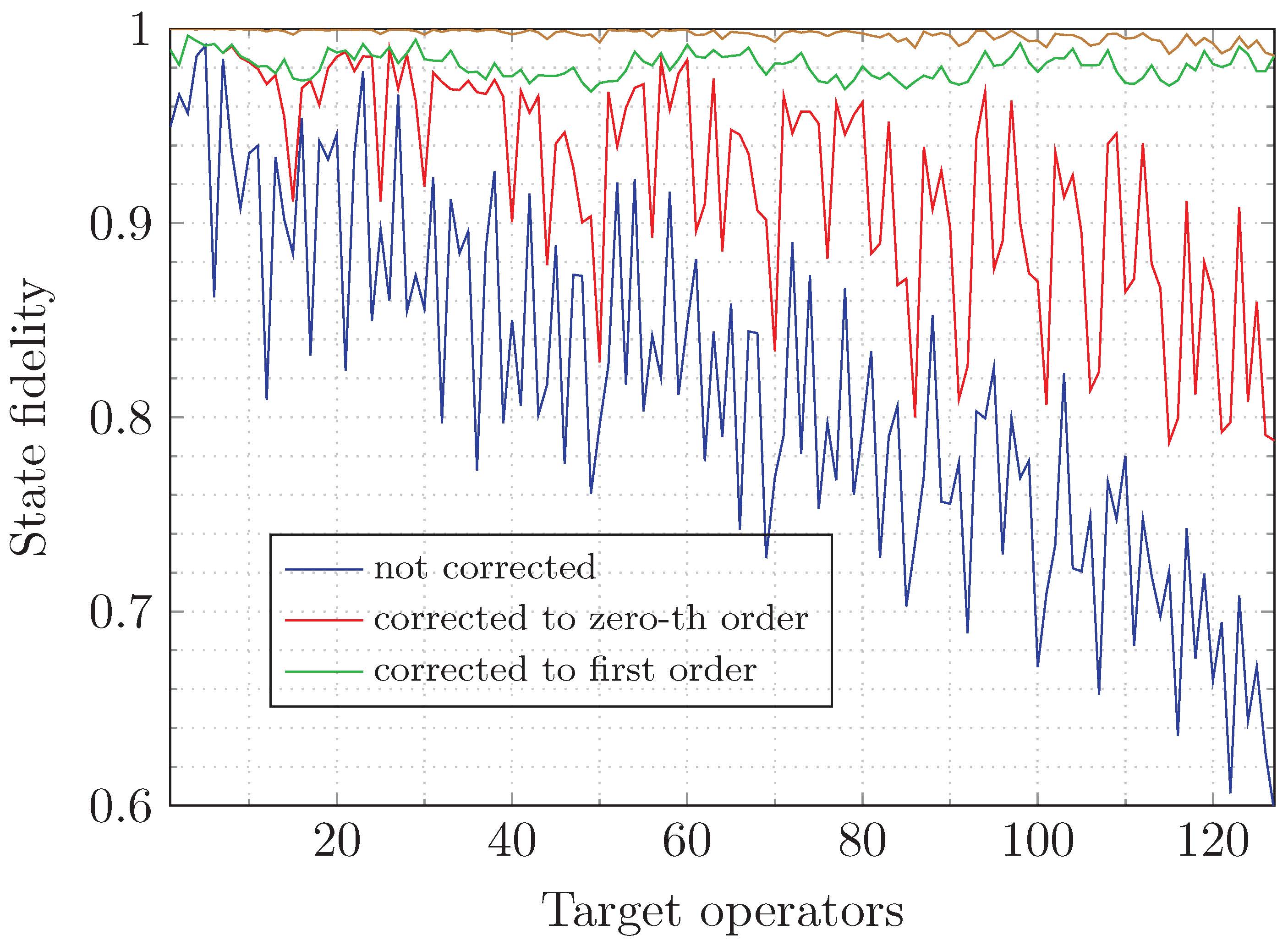}
\caption{(Color online) Simulated results of state fidelity of the $127$ single- and multi- qubit $\pi$ rotational transformations on  dichlorocyclobutane, using the selective pulse compilation method. 
For all  selective excitations, we use Gaussian pulse shapes with $T=2.2$ ms, $\sigma=T/(4\sqrt{2})$ and $\tau =20$ $\mu$s. We also plotted the numerical results under the condition of zero couplings and first-order correction for comparison (brown).}
\label{ZZZZZZZ}
\end{figure}

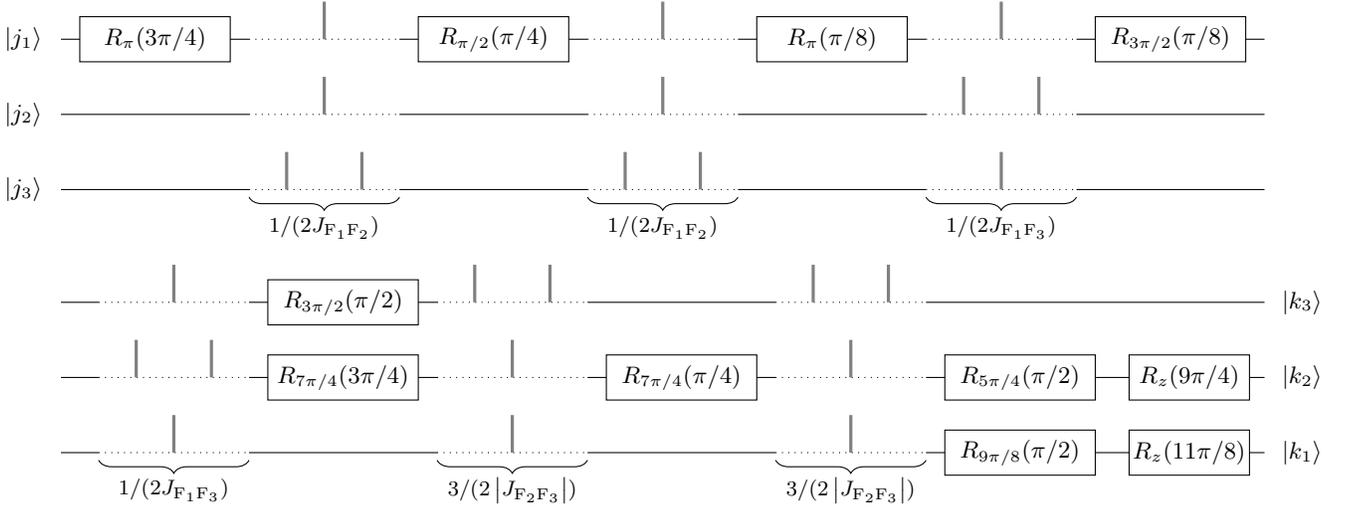
\begin{figure*}
\begin{center}

\begin{tikzpicture}

  \foreach \n in {-1,0,1} \draw  (0,\n) -- (2.5,\n);
  \foreach \n in {-1,0,1} \draw  (4.5,\n) -- (7,\n);
  \foreach \n in {-1,0,1} \draw  (9,\n) -- (11.5,\n);
  \foreach \n in {-1,0,1} \draw  (13.5,\n) -- (16,\n);

  \node at (-0.5,1) {$\left| j_1 \right\rangle$};
  \node at (-0.5,0) {$\left| j_2 \right\rangle$};
  \node at (-0.5,-1) {$\left| j_3 \right\rangle$};

  \draw  [fill=white] (1.25-1,1-0.3) rectangle (1.25+1,1+0.3);
  \node at (1.25,1) {$R_{\pi} (3\pi/4)$};

  \draw [decorate,decoration={brace,amplitude=5pt,mirror}] (2.5,-1.1) -- (4.5,-1.1);
  \node at (3.5,-1.5) {\footnotesize $1/(2 J_{\text{F}_1\text{F}_2})$};
  \foreach \n in {-1,0,1} \draw [dotted] (2.5,\n) -- (4.5,\n);
  \draw [very thick,gray] (3,-0.5)  -- (3,-1);
  \draw [very thick,gray] (3.5,0.5) -- (3.5,0) ;
  \draw [very thick,gray] (3.5,1.5) -- (3.5,1);
  \draw [very thick,gray] (4,-0.5) -- (4,-1);
  
  \draw  [fill=white] (5.75-1,1-0.3) rectangle (5.75+1,1+0.3);
  \node at (5.75,1) {$R_{\pi/2} (\pi/4)$};
  
  \draw [decorate,decoration={brace,amplitude=5pt,mirror}] (7,-1.1) -- (9,-1.1); 
  \node at (8,-1.5) {\footnotesize $1/(2 J_{\text{F}_1\text{F}_2})$}; 
  \foreach \n in {-1,0,1} \draw [dotted] (7,\n) -- (9,\n);
  \draw [very thick,gray] (7.5,-0.5) -- (7.5,-1) ;
  \draw [very thick,gray] (8,0.5) -- (8,0);
  \draw [very thick,gray] (8,1.5) -- (8,1);
  \draw [very thick,gray] (8.5,-0.5) -- (8.5,-1);
  
  \draw  [fill=white] (10.25-1,1-0.3) rectangle (10.25+1,1+0.3);
  \node at (10.25,1) {$R_{\pi} (\pi/8)$};

  \draw [decorate,decoration={brace,amplitude=5pt,mirror}] (11.5,-1.1) -- (13.5,-1.1); 
  \node at (12.5,-1.5) {\footnotesize $1/(2 J_{\text{F}_1\text{F}_3})$};   
  \foreach \n in {-1,0,1} \draw [dotted] (11.5,\n) -- (13.5,\n);
  \draw [very thick,gray] (12,0.5) -- (12,0) ;
  \draw [very thick,gray] (12.5,1.5) -- (12.5,1) ;
  \draw [very thick,gray] (12.5,-0.5) -- (12.5,-1) ;
  \draw [very thick,gray] (13,0.5) -- (13,0) ;
  \draw  [fill=white] (14.75-1,1-0.3) rectangle (14.75+1,1+0.3);
  \node at (14.75,1) {$R_{3\pi/2} (\pi/8)$};

  \foreach \n in {-2.5,-3.5,-4.5} \draw  (0,\n) -- (0.5,\n);
  \foreach \n in {-2.5,-3.5,-4.5} \draw  (2.5,\n) -- (5,\n);
  \foreach \n in {-2.5,-3.5,-4.5} \draw  (7,\n) -- (9.5,\n);
  \foreach \n in {-2.5,-3.5,-4.5} \draw  (11.5,\n) -- (16,\n);
  \node at (16.5,-2.5) {$\left| k_3 \right\rangle$};
  \node at (16.5,-3.5) {$\left| k_2 \right\rangle$};
  \node at (16.5,-4.5) {$\left| k_1 \right\rangle$};

  \draw [decorate,decoration={brace,amplitude=5pt,mirror}] (0.5,-4.6) -- (2.5,-4.6); 
  \node at (1.5,-5) {\footnotesize $1/(2 J_{\text{F}_1\text{F}_3})$};  
  \foreach \n in {-2.5,-3.5,-4.5} \draw [dotted] (0.5,\n) -- (2.5,\n);
  \draw [very thick,gray] (1,-3) -- (1,-3.5);
  \draw [very thick,gray] (1.5,-2) -- (1.5,-2.5);
  \draw [very thick,gray] (1.5,-4) -- (1.5,-4.5);
  \draw [very thick,gray] (2,-3) -- (2,-3.5);
  
  \draw  [fill=white] (3.75-1,-2.5-0.3) rectangle (3.75+1,-2.5+0.3);
  \node at (3.75,-2.5) {$R_{3\pi/2} (\pi/2)$};
  \draw  [fill=white] (3.75-1,-3.5-0.3) rectangle (3.75+1,-3.5+0.3);
  \node at (3.75,-3.5) {$R_{7\pi/4} (3\pi/4)$};
  
  \draw [decorate,decoration={brace,amplitude=5pt,mirror}] (5,-4.6) -- (7,-4.6); 
  \node at (6,-5) {\footnotesize $3/(2 \left|J_{\text{F}_2\text{F}_3}\right|)$};  
  \foreach \n in {-2.5,-3.5,-4.5} \draw [dotted] (5,\n) -- (7,\n);
  \draw [very thick,gray] (5.5,-2) -- (5.5,-2.5);
  \draw [very thick,gray] (6,-3) -- (6,-3.5);
  \draw [very thick,gray] (6,-4) -- (6,-4.5);
  \draw [very thick,gray] (6.5,-2) -- (6.5,-2.5);
  
  \draw  [fill=white] (8.25-1,-3.5-0.3) rectangle (8.25+1,-3.5+0.3);
  \node at (8.25,-3.5) {$R_{7\pi/4} (\pi/4)$};
  
  \draw [decorate,decoration={brace,amplitude=5pt,mirror}] (9.5,-4.6) -- (11.5,-4.6); 
  \node at (10.5,-5) {\footnotesize $3/(2 \left|J_{\text{F}_2\text{F}_3}\right|)$};  
  \foreach \n in {-2.5,-3.5,-4.5} \draw [dotted] (9.5,\n) -- (11.5,\n);
  \draw [very thick,gray] (10,-2) -- (10,-2.5);
  \draw [very thick,gray] (10.5,-3) -- (10.5,-3.5);
  \draw [very thick,gray] (10.5,-4) -- (10.5,-4.5);
  \draw [very thick,gray] (11,-2) -- (11,-2.5);
  
  \draw  [fill=white] (12.75-1,-3.5-0.3) rectangle (12.75+1,-3.5+0.3);
  \node at (12.75,-3.5) {$R_{5\pi/4} (\pi/2)$};
  \draw  [fill=white] (12.75-1,-4.5-0.3) rectangle (12.75+1,-4.5+0.3);
  \node at (12.75,-4.5) {$R_{9\pi/8} (\pi/2)$};
  
  \draw  [fill=white] (15-0.8,-3.5-0.3) rectangle (15+0.8,-3.5+0.3);
  \node at (15,-3.5) {$R_z (9\pi/4)$};
  \draw  [fill=white] (15-0.8,-4.5-0.3) rectangle (15+0.8,-4.5+0.3);
  \node at (15,-4.5) {$R_z (11\pi/8)$};


\end{tikzpicture}
\end{center}

%
%
%
%
%
%
%
\caption{Ideal circuit that implements the $2^3$-dimensional quantum Fourier transform on iodotrifluroethylene. It is obtained through a further reduction of the circuit  given in the main text, i.e., first $H = e^{i \pi/2} e^{-i \pi I_z} e^{i \pi/2 I_y}$, $S = e^{i \pi/4} e^{-i \pi/2 I_z}$ and $T = e^{i \pi/8} e^{-i \pi/4 I_z}$ are substituted into the sequence, then the controlled gates are decomposed into single qubit rotations and free evolutions. The vertical lines in the figure  represent ideal refocusing $\pi$ pulses, they form decoupling sequences  to allow the desired coupling to occur.  Besides, the excessive $\hat z$ rotations are all moved to the end of the network. The circuit thus constructed consists of in total $25$ operations (the last $\hat z$ rotations do not count in), including multiple-qubit rotations.
}
\label{QFT}
\end{figure*}

Fig. \ref{7qubitProfile} shows the simulated seven-site inversion profile, with each site corresponding to a resonant frequency of a carbon of labeled dichlorocyclobutanone, using and not using  the frequency shift correction scheme. Not corrected, it can be seen that the centers of the inversion profiles have shifted in frequency  and the inversion is incomplete.  Corrected, the inversion profile is almost perfect; there is very little leftover $xy$ magnetization at the seven sites.  This example demonstrates the application of our rules in constructing multi-frequency selective excitation pulse, which can  be used for  multiple-qubit gates.

Fig. \ref{ZZZZZZZ} shows the simulated final state fidelity for $\pi$ rotational state-to-state transfer problem on the labeled dichlorocyclobutanone. The initial state is assumed to be the seven-correlated spin operator $\rho_i = 2^7 I^1_z I^2_z I^3_z I^4_z I^5_z I^6_z I^7_z$, and the set of target $\pi$ operations is
\begin{equation}
\pi_1, \pi_2,  ..., \pi_1\pi_2,\pi_1\pi_3,...,   \pi_1\pi_2\pi_3\pi_4\pi_5\pi_6\pi_7,
\label{pi}
\end{equation}
which contains in total 127 elements. The state fidelity formula used for the simulation reads
\begin{equation}
\operatorname{Tr} \left[ U_\text{sim} \cdot \rho_i \cdot U^\dag_\text{sim} \cdot \left( U_\text{ideal} \cdot \rho_i \cdot U^\dag_\text{ideal} \right)^\dag \right]/2^7.
\label{statef}
\end{equation}
It can be readily seen from the figure that: (i) with only zeroth order correction, as the number of qubits being excited increases the fidelity decreases drastically and accordingly becomes more and more unacceptable; (ii)  after first-order correction the fidelity  is   improved to   the level of 0.98. 
This example demonstrates the application of our rules to perform multiple-qubit rotations.

\begin{figure}[t]
\includegraphics[width=\linewidth]{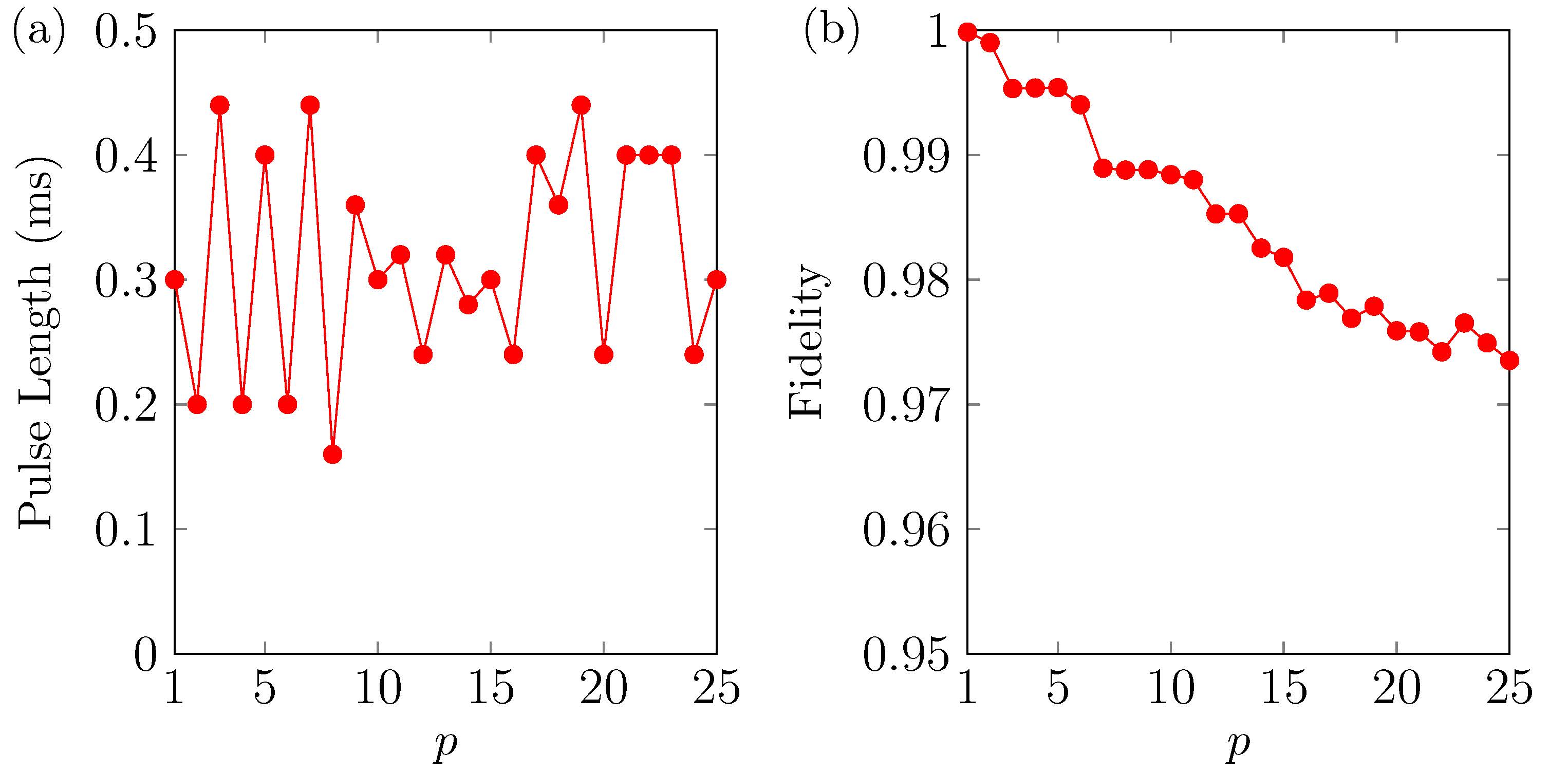}
\caption{(Color online) Simulated results of the compiled selective pulse network for the QFT circuit (Fig. \ref{QFT}). (a) The calibrated pulse length of the selective pulse used for implementing the $p$-th operation $U^p_{\text{ideal}}$. (b) The calculated fidelity between $\tilde C^p_{\text{sim}}$ and $C^p_{\text{ideal}}$.}
\label{QFTdata}
\end{figure}

Now we present an example of  compiling a selective pulse network on the three-qubit sample iodotrifluroethylene. The network is intended to  implement the $2^3$-dimensional quantum Fourier transform $\text{QFT}_{2^3}$, which is defined according to
\[ \left\vert j \right\rangle \to    \frac{1}{\sqrt{2^3}} \sum_{k=0}^{2^3-1} {e^{2\pi i jk/2^3} \left\vert k \right\rangle }.\] 
One of its circuit realizations, in terms of binary representation, is given as below \cite{NC00}
\begin{center}
\begin{tikzpicture}

  \foreach \n in {3,4,5} \draw  (0,\n) -- (6,\n);
  \node at (-0.5,5) {$\left| j_1 \right\rangle$};
  \node at (-0.5,4) {$\left| j_2 \right\rangle$};
  \node at (-0.5,3) {$\left| j_3 \right\rangle$};
  \node at (6.5,5) {$\left| k_3 \right\rangle$};
  \node at (6.5,4) {$\left| k_2 \right\rangle$};
  \node at (6.5,3) {$\left| k_1 \right\rangle$};

  \draw  [fill=white] (0.5-0.3,5-0.3) rectangle (0.5+0.3,5+0.3);
  \node at (0.5,5) {$H$};
  \draw  [fill=white] (1.5-0.3,5-0.3) rectangle (1.5+0.3,5+0.3);
  \node at (1.5,5) {$S$};
  \draw [fill=black] (1.5,4) circle [radius=0.05];
  \draw (1.5,4) to (1.5,5-0.3);
  \draw  [fill=white] (2.5-0.3,5-0.3) rectangle (2.5+0.3,5+0.3);
  \node at (2.5,5) {$T$};
  \draw [fill=black] (2.5,3) circle [radius=0.05];
  \draw (2.5,3) to (2.5,5-0.3);
  \draw  [fill=white] (3.5-0.3,4-0.3) rectangle (3.5+0.3,4+0.3);
  \node at (3.5,4) {$H$};
  \draw  [fill=white] (4.5-0.3,4-0.3) rectangle (4.5+0.3,4+0.3);
  \node at (4.5,4) {$S$};
  \draw [fill=black] (4.5,3) circle [radius=0.05];
  \draw (4.5,3) to (4.5,4-0.3);
  \draw  [fill=white] (5.5-0.3,3-0.3) rectangle (5.5+0.3,3+0.3);
  \node at (5.5,3) {$H$};
\end{tikzpicture}
\end{center}
where  $H$, $S$ and $T$ are Hadamard gate, phase gate and $\pi/8$ gate respectively. First, we  convert  the above circuit into the same form as Eq. (\ref{Cideal}). As $\hat z$ rotations do not require real physical pulses, we convert as many transverse rotations as possible into them.
Fig. \ref{QFT} shows the resulting circuit. This  ideal circuit exclusive of the final $\hat z$ rotations can be written as $\left\{t_p,U^p_{\text{ideal}}\right\}$ with $p=1,2,...,25$, note that all these operations are separated in time. Let $C^p_{\text{ideal}}$ denote the corresponding operation to the truncated circuit from the start to time $t_p$, that is, \[C^p_{\text{ideal}} =  U^p_{\text{ideal}}  \exp({-i H_S (t_p-t_{p-1})}) \cdots U^1_{\text{ideal}}.\] Thus, $C^{25}_{\text{ideal}}$ equals to the Fourier transform $\text{QFT}_{2^3}$, up to $\hat z$ rotations and  permutation of basis.
Now we replace  the ideal rotational operations in the circuit by selective pulses. Not surprisingly the  imperfections soon accumulate, causing severe  reduction of control accuracy. With application of the compilation method as described in the previous section, there is, the evolution generated by the pulse network up to time $t_p$: $C^p_{\text{sim}}  \approx U^{\hat z}_{\text{post}} \cdot \tilde C^p_{\text{sim}}$, where $U^{\hat z}_{\text{post}}$ represents post $\hat z$ errors.
As $U^{\hat z}_{\text{post}}$ is irrelevant, we just compare $\tilde C^p_{\text{sim}}$ with $C^p_{\text{ideal}}$.
Fig. \ref{QFTdata} gives the results, from which we  find that the error rate is greatly decreased as expected. The compilation procedure is quite efficient that it finishes within just seconds. From the data, the transform produced by the compiled pulse sequence is of  fidelity about 0.9735 compared with  $\text{QFT}_{2^3}$ ignoring   observation phase adjustment. Although finding a pulse of almost perfect fidelity for $\text{QFT}_{2^3}$ can be easily done through optimal searching, what we   demonstrate here is a scalable way of doing quantum computation with taking selective pulse as  building elements for the control.   

\section{Summary}
We have developed an effective and simple scheme to adjust the pulses in a selective pulse network to improve quantum control. Computing the  corrections for each pulse shape is straightforward and no optimization is involved thus the compilation procedure is quite efficient. The framework developed in this paper is in several aspects different from Ref. \cite{Ryan08}, among which most distinctly: (i) an ideal circuit is given in advance, based on which our pulse network compilation is done, i.e., we do not use an optimization programme (sequence compiler) to adjust the time delays between the selective pulses. Therefore it gains   simplicity of compilation procedure at the cost of lowering the control accuracy; (ii) we have taken into account multiple-qubit operations, which  requires us to make  analysis of the first order average Hamiltonian.
We have  confirmed, with a series of numerical tests, that our method can  be applied to increase control fidelity.
The control techniques developed here could be easily incorporated into existing pulse design schemes.  We expect them to be applied to larger molecules and be extended to other experimental systems.

\section{Acknowledgments}
Jun Li completed part of this study while visiting the Institute for Quantum Computing at the University of Waterloo and acknowledges for what he has learned from the pulse design methods and  programmes developed in Raymond Laflamme's research group. This work is supported by the National Basic Research Program of China (973 Program, Grant No. 2014CB921403), the National Key Research and Development Program (Grant No. 2016YFA0301201), the Foundation for Innovative Research Groups of the National Natural Science Foundation of China (Grant No. 11421063),  the  Major Program of the National Natural Science Foundation of China (Grant No. 11534002), the State Key Development Program for Basic Research of China (Grant Nos. 2014CB848700 and 2013CB921800), the National Science Fund for Distinguished Young Scholars (Grant No. 11425523), and the National Natural Science Foundation of China (Grant No. 11375167).

\appendix


\begin{widetext}

\section{First Order Averaged Hamiltonian}
The first order averaged Hamiltonian has the following expression
\begin{align}
\mathcal{H}^{(1)}_1   = {} &  \frac{-i}{2T} \left( \sum\limits_{m_2=1}^M \sum\limits_{m_1=1}^{m_2}  \int_{(m_2-1)\tau}^{m_2\tau}  \int_{(m_1-1)\tau}^{m_1\tau} - \sum\limits_{m_2=1}^M \sum\limits_{m_1=m_2}  \int_{(m_2-1)\tau}^{m_2\tau}  \int_{t_2}^{m_1 \tau}  \right)  \nonumber \\
{} & \left[ \left[   \sum\limits_{k_2=1}^n \sum\limits_{j_2=1}^n {\theta'_{j_2} {G}_{j_2} [m_2] I^{k_2}_{\Omega'_{j_2} m_2\tau -\Omega_{k_2} t_2 + \varphi'_{j_2}} },   \sum\limits_{k_1=1}^n \sum\limits_{j_1=1}^n {\theta'_{j_1} {G}_{j_1} [m_1] I^{k_1}_{\Omega'_{j_1} m_1\tau - \Omega_{k_1} t_1 + \varphi'_{j_1}} }  \right] dt_1 dt_2 \right]  \nonumber \\
= {} & \frac{1}{2T}   \sum\limits_{k=1}^n \sum\limits_{j_2=1}^n \sum\limits_{j_1=1}^n  \left( \sum\limits_{m_2=1}^M \sum\limits_{m_1=1}^{m_2}  \int_{(m_2-1)\tau}^{m_2\tau}  \int_{(m_1-1)\tau}^{m_1\tau} - \sum\limits_{m_2=1}^M \sum\limits_{m_1=m_2}  \int_{(m_2-1)\tau}^{m_2\tau}  \int_{t_2}^{m_1 \tau} \right) \nonumber \\
{} &   \left[ \theta'_{j_2}\theta'_{j_1} {G}_{j_2} [m_2] {G}_{j_1} [m_1]  \sin \left((\Omega'_{j_1} m_1 - \Omega'_{j_2} m_2)\tau - \Omega_{k} (t_1 -  t_2) + \varphi'_{j_1} - \varphi'_{j_2} \right) I^{k}_z  dt_1 dt_2  \right] \nonumber
\end{align}
There is
\begin{align}
\mathcal{H}^{(1)}_1   
= {} &  \frac{1}{2T}   \sum\limits_{k=1}^n \sum\limits_{j_2=1}^n \sum\limits_{j_1=1}^n  \sum\limits_{m_2=1}^M \sum\limits_{m_1=1}^{m_2} \nonumber \\
{} &  \left[ \theta'_{j_2}\theta'_{j_1} {G}_{j_2} [m_2] {G}_{j_1} [m_1] \tau^2 \frac{\sin^2(\Omega_k\tau/2)}{(\Omega_k \tau/2)^2}  \sin \left(  (\Omega'_{j_1} - \Omega_{k}) m_1\tau -(\Omega'_{j_2} - \Omega_{k}) m_2\tau + \varphi'_{j_1} - \varphi'_{j_2} \right)   I^k_z  \right]   \nonumber \\
{} & - \frac{1}{2T}   \sum\limits_{k=1}^n \sum\limits_{j_2=1}^n \sum\limits_{j_1=1}^n  \sum\limits_{m=1}^M \left[ \theta'_{j_2}\theta'_{j_1} {G}_{j_2} [m] {G}_{j_1} [m]  \frac{\tau}{\Omega_k}  \right.\nonumber \\
{} & \left. \left( \frac{\sin(\Omega_k\tau/2)}{\Omega_k \tau/2} \cos( (\Omega'_{j_2} - \Omega'_{j_1}) m\tau - \Omega_k \tau/2 + \varphi'_{j_1} - \varphi'_{j_2})  - \cos( (\Omega'_{j_2} - \Omega'_{j_1}) m\tau + \varphi'_{j_1} - \varphi'_{j_2})  \right) I^k_z \right]
\label{BSshift}
\end{align}
Here, again we would retain only those terms for which $j_1 = j_2$
\begin{align}
\mathcal{H}^{(1)}_1   =  {} &  \frac{1}{2T}   \sum\limits_{k=1}^n \sum\limits_{j=1}^n  \sum\limits_{m_2=1}^M {\left[ {\theta'_{j}}^2 {G}_{j} [m_2] \tau^2  \frac{\sin^2(\Omega_k\tau/2)}{(\Omega_k \tau/2)^2} \left(  \sum\limits_{m_1=1}^{m_2} { {G}_{j} [m_1]  \sin \left((\Omega'_{j} - \Omega_{k}) (m_1-m_2)\tau \right)} \right)  I^k_z  \right]}  \nonumber  \\
{} & - \frac{1}{2T}   \sum\limits_{k=1}^n \sum\limits_{j=1}^n   \sum\limits_{m=1}^M \left[ {\theta'_{j}}^2 {G}_j[m]^2 \frac{\tau}{\Omega_k} \left(\frac{\sin(\Omega_k\tau)}{\Omega_k \tau} - 1 \right) I^k_z \right].
\end{align}
The latter summation is negligible. We divide $\mathcal{H}^{(1)}_1$ into two parts: $\mathcal{H}^{(1,1)}_1$ for $j=k$ and $\mathcal{H}^{(1,2)}_1$ for $j \ne k$.
For $j=k$:
\begin{align}
\mathcal{H}^{(1,1)}_1   =  {} &  \frac{1}{2T}   \sum\limits_{k=1}^n {\left( {\theta'_{k}}^2 \tau^2 \frac{\sin^2(\Omega_k\tau/2)}{(\Omega_k \tau/2)^2} \sum\limits_{m_2=1}^M \sum\limits_{m_1=1}^{m_2} {  {G}_{k} [m_2]   { {G}_{k} [m_1]  \sin \left(\Delta \Omega_{k} (m_1-m_2)\tau \right)}  } \right) I^k_z } \nonumber \\
\approx {} &  \frac{1}{2T} \sum\limits_{k=1}^n {\left(\Delta \Omega_{k} {\theta'_{k}}^2 \tau^3 \frac{\sin^2(\Omega_k\tau/2)}{(\Omega_k \tau/2)^2} \sum\limits_{m_2=1}^M \sum\limits_{m_1=1}^{m_2} {  {G}_{k} [m_2]   { {G}_{k} [m_1]   (m_1-m_2) } } \right) I^k_z } \nonumber  \\
= {} &   \sum\limits_{k=1}^n {\eta_k \Delta \Omega_{k} I^k_z }. 
\label{H111}
\end{align}
\end{widetext}
For $j \ne k$,  we would perform some perturbative techniques, that is, Fourier coefficient asymptotics. Basically, we need the following lemma \cite{Hinch91}:
\begin{lemma}
If $f(t) \in C^\infty[0,T]$, then there is the asymptotic expansion
\begin{align}
\int_0^T {f(t) e^{i \omega t} dt} = {} & \sum\limits_{n=1}^{N} {\frac{(-1)^{n-1}}{(i\omega)^n} \left.\left[ f^{(n-1)}(t) e^{i\omega t} \right] \right\vert_0^T}  \nonumber \\
{} & + \frac{(-1)^{N}}{(i\omega)^{N+1}} \int_0^T {f^{(N)}(t) e^{i\omega t}dt}. \nonumber
\end{align}
\end{lemma}
This is readily established, using integration by parts iteratively. Therefore we have
\begin{equation}
\int_0^T {f(t) \sin(\omega t) dt} = -\frac{1}{\omega} \left.\left\{ f(t) \cos(\omega t) \right\} \right\vert_0^T + O(\omega^{-2}). \nonumber
\end{equation}
As our expression is a summation rather than an integral, we thus uses the analogue formula, that is, summation by parts (known as the Abel's lemma).

In \cite{Warren84,EB90,ZG98},  approximate expression for the transient Bloch-Siegert shifts were given. Following their approaches, we made a derivation for  the general case at the appendix, the result is
\begin{equation}
\Delta \Omega_k^{\text{BS}} \approx   \sum\limits_{j \ne k}^n {\left( - \frac{\sin^2(\Omega'_k\tau/2)}{(\Omega'_k \tau/2)^2}  \frac{ {\theta'_{j}}^2 \mathbb{E}({G}^2_j) }{2(\Omega'_{j} - \Omega'_{k})} \right) I^k_z}, 
\end{equation}
It can be readily verified that in the case of rectangle wave irradiation, it reduces to the normal Bloch-Siegert shift. 

The estimation proceeds as follows
\begin{align}
{} & \sum\limits_{m_1=1}^{m_2} { {G}_{j} [m_1]  \sin \left((\Omega'_{j} - \Omega'_{k}) (m_1-m_2)\tau   \right) } \nonumber \\
\approx {} & \left. \left\{ {G}_{j} [m_1] \frac{- \cos \left((\Omega'_{j} - \Omega'_{k}) (m_1-m_2) \tau  \right)}{(\Omega'_{j} - \Omega'_{k})\tau } \right\}\right\vert_{m_1=1}^{m_2}   \nonumber \\
\approx {} & \frac{ {G}_{j} [m_2] }{(\Omega'_{j} - \Omega'_{k})\tau } (-1 + \cos \left((\Omega'_{j} - \Omega'_{k}) (1-m_2) \tau  \right)).  \nonumber
\end{align}
Substitute the above equation into $\mathcal{H}^{(1,2)}_1$ and omitting the fast oscillating sum, we finally get
\begin{equation}
\mathcal{H}^{(1,2)}_1 \approx \sum\limits_{k=1}^n \sum\limits_{j \ne k}^n {\left( - \frac{\sin^2(\Omega'_k\tau/2)}{(\Omega'_k \tau/2)^2}  \frac{ {\theta'_{j}}^2 \mathbb{E}({G}^2_j) }{2(\Omega'_{j} - \Omega'_{k})} \right) I^k_z}, 
\label{H112}
\end{equation}
where we have defined
\begin{equation}
\mathbb{E}({G}^2_j) = \frac{1}{T} \sum\limits_{m=1}^M  { {G}^2_{j} [m] \tau}.
\label{E}
\end{equation}

\end{document}